\documentclass[11pt]{article}
\usepackage[utf8]{inputenc}
\usepackage[T1]{fontenc}
\usepackage[margin=1in]{geometry}
\usepackage{graphicx}
\usepackage{subcaption}
\usepackage{setspace}
\usepackage{amsmath}
\usepackage[authoryear, round]{natbib}

\title{Modeling and Inferring Metacommunity Dynamics with Maximum Caliber}

\author{Zachary Jackson\textsuperscript{*1}, Mathew A. Leibold\textsuperscript{2}, Robert D. Holt\textsuperscript{2}, BingKan Xue\textsuperscript{1}}
\date{\textsuperscript{1} Department of Physics, \textsuperscript{2} Department of Biology,\\
University of Florida, Gainesville, FL, United States}

\begin{document}

\maketitle

\begin{abstract}
A major challenge for community ecology is using spatio-temporal data to infer parameters of dynamical models without conducting laborious experiments. We present a novel framework from statistical physics -- Maximum Caliber -- to characterize the temporal dynamics of complex ecological systems in spatially extended landscapes and infer parameters from empirical data. As an extension of Maximum Entropy modeling, Maximum Caliber aims at modeling the probability of possible trajectories of a stochastic system, rather than focusing on system states. We demonstrate the ability of the Maximum Caliber framework to capture ecological processes ranging from near- to far from- equilibrium, using an array of species interaction motifs including random interactions, apparent competition, intraguild predation, and non-transitive competition, along with dispersal among multiple patches. For spatio-temporal data of species occupancy in a metacommunity, the parameters of a Maximum Caliber model can be estimated through a simple logistic regression to reveal migration rates between patches, interactions between species, and local environmental suitabilities. We test the accuracy of the method over a range of system sizes and time periods, and find that these parameters can be estimated without bias. We introduce ``entropy production'' as a measure of irreversibility in system dynamics, and use ``pseudo-$R^2$'' to characterize predictability of future states. We show that our model can predict the dynamics of metacommunities that are far from equilibrium. The capacity to estimate basic parameters of dynamical metacommunity models from spatio-temporal data represents an important breakthrough for the study of metacommunities with application to practical problems in conservation and restoration ecology.
\end{abstract}

\clearpage

\section*{Introduction}

The basic processes that drive community changes are relatively simple \citep{vellend:2010, vellend:2016}, consisting of dispersal, ``abiotic filtering'' (density independent selection among species), ``biotic filtering'' (density/frequency dependent selection), stochasticity, and the generation of ``novelty'' (especially speciation).  However, the resulting community patterns over the long term arise from complex dynamics that reflect the wide-spread ``entanglement'' of these causal processes \citep{darwin:1859, kefi:2016}. It is possible, using various theoretical approaches, to develop hypotheses and derive predictions from particular model assumptions and parameterizations. However, the inverse problem, using data to infer the processes that lead to these patterns, is much more difficult \citep{leibold:2025}.  This is because different models often make convergent predictions about steady states, and also because disentangling causal effects is not always possible using current approaches based on data about spatial distributions alone.  

Most prior work to derive predictions about community patterns has used population models based on the classic Lotka-Volterra formulation of species interactions \citep{lotka:1925, volterra:1928}, expanded to multi-species communities \citep{may:1972,novak:2016} and multiple patches and trophic levels \citep{gravel:2016}.  These models are difficult to solve because they are non-linear in their formulation, especially when dispersal among communities occurs.  A simpler modeling approach ignores variation in abundances and focuses on occupancy (presence or absence) instead.  Such models examine how colonizations and extinctions affect community patterns \citep{levins:1971, horn:1972, holt:1997:2, leibold:2022:2}, typically focusing on the steady state of species distributions.  Here we use occupancy models, but focus on spatio-temporal dynamics, both near steady state and far from equilibrium, to infer causal relations in metacommunities.

Because species occupancy is a result of stochastic ecological processes, a ``steady state'' means that the \emph{probability} of a species occupying a given patch is stationary over time. It does not mean that the species has to be always present or always absent at the patch, only that its frequency of occupancy (over short periods) stops changing (over the long term). In contrast, ``transient dynamics'' are those during which the \emph{probability} of some species occupying a given patch changes over time. The dynamics of a system can be called ``at equilibrium'', or ``time reversible'', if the probability of any state transition is equal to that of the reverse transition. We note that steady states are not necessarily at equilibrium, since there can be nonvanishing probability currents, such as in assembly cycles; transient dynamics are even farther from equilibrium, as the system follows trajectories that are strongly directional.

With changes in data pipelines and development of statistical methods, there is an increasing access to spatio-temporal data in ecology \citep{hartig:2024}, permitting the use of methods that take advantage of temporal trajectories to better identify causal relations.  Purely spatial models conventionally utilize a basic sample dataset (lists of species present at a particular patch at a particular time) that characterize the ``states'' of communities.  However, an alternative approach is to focus on changes in these states from one time point to the next, consisting of ``transitions''. An accumulation of transitions over multiple time steps forms ``trajectories'', which can describe both the equilibrium outcome of community dynamics and transient, far-from equilibrium behaviors.  This is the basic approach we explore here.

We introduce a method from statistical physics that squarely addresses such trajectories, known as Maximum Caliber \citep{jaynes:1980,stock:2008}. It has been applied to stochastic processes in a variety of research areas, including thermal and chemical dynamics, protein folding, neural firing, network traffic flows, and more \citep{presse:2013, ghosh:2020, wan:2016, firman:2017, mora:2015}. Here we apply Maximum Caliber to metacommunity ecology and find that this approach of modeling the behavior of metacommunities allows a straightforward way to infer parameters that can be interpreted as basic ecological processes.  This method thus opens up previously unavailable, but critically important, linkages between observable patterns in metacommunities and the underlying processes that are responsible for them.

We first derive a general mathematical framework for using Maximum Caliber to study community assembly and dynamics in metacommunities.  We then apply this approach to a representative, albeit simple, set of models that involve three species in three discrete patches, ranging from fully random to highly deterministic interactions, so as to demonstrate the utility of Maximum Caliber in describing and analyzing metacommunity dynamics.  We quantify the irreversibility of their dynamics by calculating the average rate of entropy production \citep{wang:2020}. We then show that Maximum Caliber can make accurate predictions of data, as measured by the pseudo-$R^2$ value, especially for systems far from equilibrium.  An important advantage of our approach is that we can make unbiased estimation of model parameters from observations using simple logistic regression. Of course, the accuracy of the model crafted using Maximum Caliber depends on data quantity and quality, but we expect that realistic results can be obtained with the kind of data increasingly available to ecologists.

\begin{figure*}[!ht]
\centering
\includegraphics[width=\textwidth]{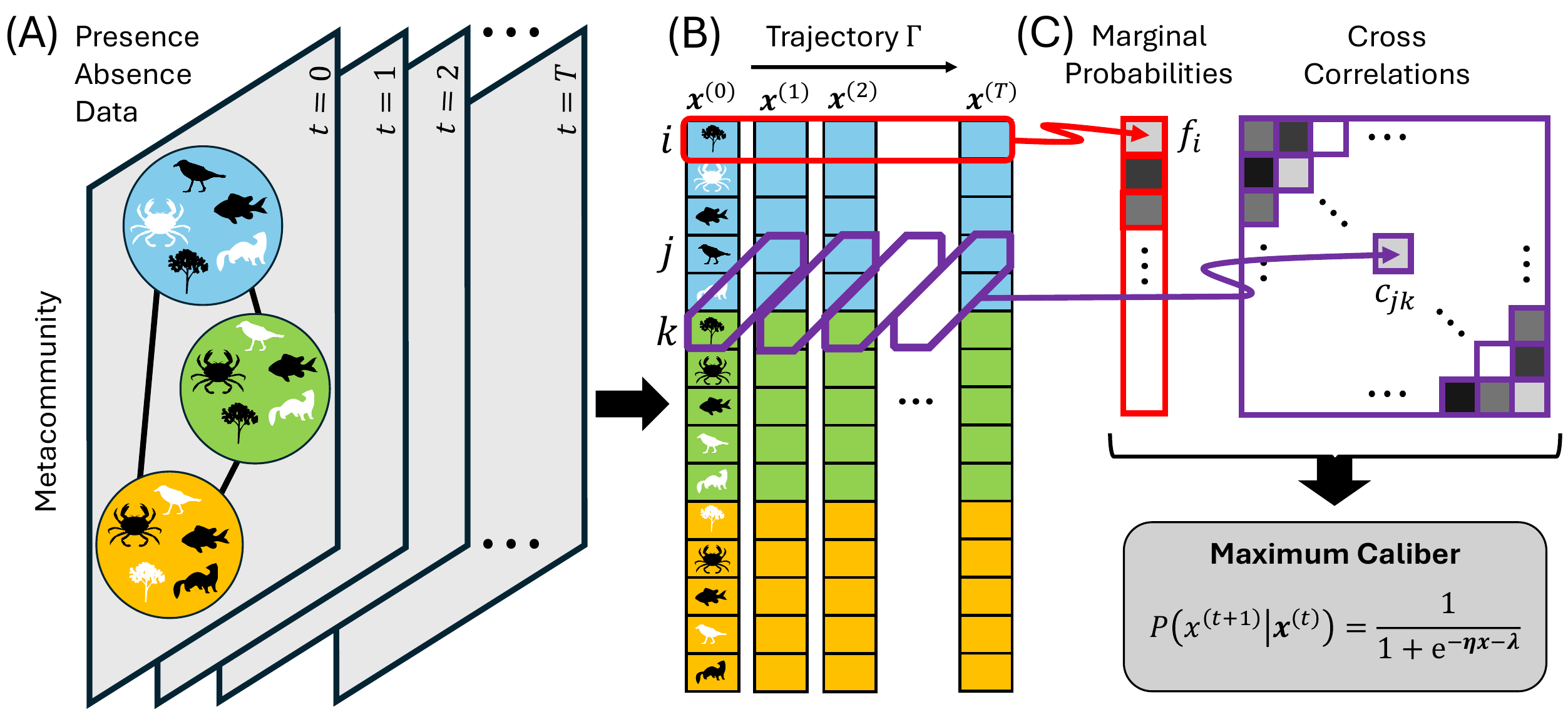}
\caption{\small The Maximum Caliber framework for modeling spatio-temporal dynamics of a metacommunity. (A) We consider ecological data in the format of the presence or absence of each species from a species pool, measured in every patch of the metacommunity over many time points. (B) The presence/absence data of every species-patch pair (labeled by $i, j, k$, etc.) at a given time point $t$ is represented by a binary vector $\mathbf{x}^{(t)}$, where each entry $x^{(t)}_i$ is $1$ if the species is present at the corresponding patch or 0 otherwise. The time series of the vector $\mathbf{x}^{(t)}$ forms a trajectory $\Gamma$ in the state space. (C) From the time series data we can calculate the marginal probability $f_i$ of each species-patch pair $i$, which represents the frequency of the species occupying the corresponding patch, and the cross-correlation $c_{jk}$ between a species-patch pair $j$ at time $t+1$ and another species-patch pair $k$ at time $t$. These marginal probabilities and cross-correlations will be used to build a Maximum Caliber model of the metacommunity dynamics (Box~1), which describes the probability of system trajectories $\Gamma$ and predicts the future state given the history of the metacommunity.}
\label{fig:scheme}
\end{figure*}

\section*{Methods}

We consider a simplified view of metacommunity dynamics (Fig.~\ref{fig:scheme}) that has a venerable history in community ecology \citep{levins:1971, horn:1972, levin:1974, hastings:1980, hanski:1983, holt:1997, leibold:2015, leibold:2022:2}. This perspective assumes that there are different discrete locations (``patches''), which can potentially be occupied by species that are subsets of a defined regional ``biota''.  These patches can differ in local environmental conditions that define the ``abiotic selection'' affecting occupancy by different species, modeled as the intrinsic capacity of species to colonize the patch (assuming the absence of other species), and the intrinsic vulnerability of species to go extinct there (also assuming the absence of other species).  These intrinsic capacities are modified by the presence of other species due to species interactions, which can alter either the colonization probability or the extinction probability of any given species relative to what would be the case in similar patches not occupied by these other species.  We also assume that dispersal plays a role, either by connectivity among patches in the metacommunity (perhaps scaled by physical distance), or due to a more diffuse (typically much smaller) ``global dispersal'' from external sources that is uniform across all patches.  Colonization and extinction processes are all modeled as probabilities, so that there is a substantial amount of stochasticity in the model dynamics, the magnitude of which depends on the parameter values.  

These processes when iterated generate spatio-temporal lists of species on different patches. These datasets can be represented as state vectors with elements, either 0 or 1, describing the occupancy of patches by individual species at any given time $t$.  However, the data can also be viewed as describing transitions among such states from one time point to the next. Within a single patch with a potential pool of $N$ species, the state of the local community at a given time can be one of $2^N$ possible states, depending on whether each species is present or not, and the possible number of transitions is $2^N \times 2^N$. The dynamics of the community can therefore be described by repeated transitions of these states over time. If we have $N$ species distributed among $M$ patches, the number of states that describe the entire metacommunity is $2^{MN}$, and the number of possible transitions over one time step is $2^{MN} \times 2^{MN}$. Deriving these probabilities by brute force is consequently a daunting task, except for cases with very few species and very few patches.

One approach in this situation is to estimate the marginal probability of each state and then use that probability to predict future states, regardless of the likelihood of state transitions. That is, we can assign a probability to each possible state of the system according to its observed frequency, regardless of the temporal context (Fig.~\ref{fig:scheme}B top panel). Maximum Entropy  \citep{jaynes:1957, shipley:2010, harte:2014} is one such modeling scheme grounded in this approach that parameterizes the marginal probabilities for the states of a system, while attempting to use as few additional assumptions as possible. Maximum Entropy can successfully describe and predict many features of systems that are at or near equilibrium, but it is less informative otherwise.  Yet, ecological systems are often far from equilibrium \citep{deangelis:1987, hastings:1993} and can have complex dynamics that are not captured by marginal probabilities.

To obtain a more effective model that accounts for non-equilibrium spatio-temporal community dynamics, we turn to the Maximum Caliber framework that has been developed for modeling random processes in physics \citep{jaynes:1980, presse:2013, ghosh:2020}. Maximum Caliber is a modeling scheme similar to Maximum Entropy, but it considers probabilities of possible trajectories of the system rather than probabilities of states. In the Maximum Caliber approach, predicting the future state of a system can be thought of as finding the most likely trajectory of the system in the state space over time. By focusing on the system trajectories rather than states, Maximum Caliber can capture dynamics of systems far from equilibrium as well as near or at equilibrium.

\subsection*{Maximum Caliber}

The aim of this approach is to develop a model for system trajectories that is consistent with a given set of observational constraints, but is otherwise as unbiased as possible. If we choose the constraints to be the temporal correlations between consecutive states, then Maximum Caliber can generate a model parameterized by a matrix of effective species interactions, whose entries are interpretable as the rates of basic biological processes governing transitions among states.  These parameter values are inferable by a simple logistic regression, providing direct inference of model parameters from the data in a way that has not been possible for the classic Lotka-Volterra model.

We start with a single patch subject to colonization by an external pool of species, labeled by $i = 1, \cdots, N$. We then imagine that we have at hand the list of all species present at different time points. At each time $t$, the species list can be represented by a set of $N$ binary variables, $x_i = 1$ or $0$, according to whether species $i$ is present or absent. The state of the system is thus represented by an $N$-dimensional binary vector $\mathbf{x}$. The data from all time points then forms a time series of such binary vectors, i.e., a trajectory in $N$-dimensional space. We will label the community state at time $t$ by $\mathbf{x}^{(t)}$, and denote a trajectory in the state space by $\Gamma = \{ \mathbf{x}^{(0)}, \mathbf{x}^{(1)}, \mathbf{x}^{(2)}, \cdots, \mathbf{x}^{(T)} \}$, where $T$ is the length of the time series. 

We seek to model the probability of different trajectories, $P(\Gamma)$, that are consistent with certain statistical properties of the observed data. We will focus on the frequency of occupancy of each species, $f_i$, and the cross-correlation between pairs of species at consecutive time points, $c_{ij}$. We thus look for a model of $P(\Gamma)$ that satisfies these constraints but is otherwise as unbiased as possible. This is done by maximizing the entropy of the distribution $P(\Gamma)$, under the constraints that the expected values of $f_i$ and $c_{ij}$ match the observed values from the data (see Box 1).

\begin{figure*}[!ht]
\centering
\fbox{
\begin{minipage}{\textwidth}
\section*{Box 1: Maximum Caliber Modeling}
Here is a brief description of standard derivation of a Maximum Caliber model \citep{presse:2013}.
We start from the entropy of the probability distribution $P(\Gamma)$ of all possible trajectories:
\begin{equation}
\mathcal{H} = - \sum_\Gamma P(\Gamma) \log P(\Gamma) \,.
\end{equation}
We maximize the entropy under constraints on the statistics of the trajectories, here chosen to be the occupancy frequency of each species, $f_i$, and the cross-correlation between pairs of species at consecutive time points, $c_{ij}$. For a given trajectory $\Gamma$, these quantities can be calculated as
\begin{align}
f_i (\Gamma) = \frac{1}{T} \sum_{t=0}^T x_i^{(t)}
\quad \text{and} \quad
c_{ij} (\Gamma) = \frac{1}{T} \sum_{t=0}^{T-1} x_i^{(t+1)} x_j^{(t)} \,.
\label{eq:fc_obs}
\end{align}
Suppose $f_i^\text{obs}$ and $c_{ij}^\text{obs}$ are the values calculated from the observed data. We assume that these are the expected values of $f_i$ and $c_{ij}$ according to the probability $P(\Gamma)$, i.e.,
\begin{align}
\sum_{\Gamma} P(\Gamma) f_i (\Gamma) = f_{i}^\text{obs} 
\quad \text{and} \quad
\sum_{\Gamma} P(\Gamma) c_{ij}(\Gamma) = c_{ij}^\text{obs} \,.
\label{eq:fC_con}
\end{align}
Introduce Lagrange multipliers to impose these constraints on $P(\Gamma)$, the function to be maximized becomes
\begin{align} \label{eq:Lagrange}
\mathcal{L} =& - \sum_{\Gamma} P({\Gamma}) \log P(\Gamma) + \sum_{ij} \tilde{\eta}_{ij} \left( \sum_{\Gamma} P(\Gamma) c_{ij}(\Gamma) - c_{ij}^\textrm{obs} \right)\nonumber \\ &+ \sum_{i} \tilde{\lambda}_{i} \left( \sum_{\Gamma} P(\Gamma) f_i(\Gamma) - f_{i}^\textrm{obs} \right) + \mu \left( \sum_\Gamma P(\Gamma) - 1 \right) .
\end{align}
Among the Lagrange multipliers, $\tilde{\lambda}_i$ and $\tilde{\eta}_{ij}$ constrain the frequencies and cross-correlations, and $\mu$ ensures the probabilities sum to 1. Maximizing this function leads to the solution in Eq.~[\ref{eq:cond-prob-1}], where the parameters $\lambda_{i}$ and $\eta_{ij}$ are derived from $\tilde{\lambda}_i$ and $\tilde{\eta}_{ij}$ (see more details in Materials and Methods). The values of these parameters are to be determined by solving the constraint equations [\ref{eq:fc_obs}]; due to the form of Eq.~[\ref{eq:cond-prob-1}], we can approximately infer the parameters using logistic regression (see Supplementary Text for a comparison between parameter inference in Maximum Caliber and logistic regression).
\end{minipage}
}
\end{figure*}

The Maximum Caliber solution takes the form:
\begin{equation*}
P(\Gamma) = \prod_t \prod_i P \big( x_i^{(t+1)} \big| \mathbf{x}^{(t)} \big),
\end{equation*}
where
\begin{equation}
P \left( x_i^{(t+1)} \!=\! 1 \Big| \mathbf{x}^{(t)} \right) = \frac{1}{1 + \exp \left\{ -\sum_j \eta_{ij} \, x_j^{(t)} - \lambda_i \right\}} \,.
\label{eq:cond-prob-1}
\end{equation}
This implies that the presence or absence of each species in the subsequent time point is conditional only on the community composition in the previous time point. Therefore, this is a special case of a Markov model --- but instead of being parameterized by a $2^N \times 2^N$ transition matrix $P \big( \mathbf{x}^{(t+1)} \big| \mathbf{x}^{(t)} \big)$, our model is parameterized by a much smaller $N \times N$ matrix $\eta_{ij}$ and a $N$-dimensional vector $\lambda_i$. Here the parameter $\eta_{ij}$ represents the effect of species $j$ on the persistence of species $i$ due to species interactions (including effects of species \textit{i} on itself that include intrinsic relations to the abiotic environment), and the parameter $\lambda_i$ that is related to the colonization of the patch by species $i$ from an external pool.

Since the conditional probability in Eq.~(\ref{eq:cond-prob-1}) takes the form of a logistic function, the parameters $\eta_{ij}$ and $\lambda_i$ can be estimated from a series of logistic regressions, one for each species $i$. Each time step provides a data point with $\mathbf{x}^{(t)}$ (the presence/absence of every species at time $t$) as the explanatory variables and $x_i^{(t+1)}$ (the presence/absence of species $i$ at time $t+1$) as the outcome variable. The $\eta_{ij}$'s are the regression coefficients and $\lambda_i$'s are the intercepts.

We can expand our model from characterizing a single patch to one that includes multiple patches interconnected by dispersal. That is, species can colonize individual patches not only from an external pool, but also from other patches. To describe such a metacommunity, the label $i$ now represents a species-patch combination, so that $x^{(t)}_{i}$ represents whether a particular species is present in a given patch at a given time $t$. Different entries in the $\eta_{ij}$ parameters represent either interactions between species within a patch, or rates of species-specific migration between patches (both can be asymmetric). As illustrated in Fig.~\ref{fig:motifs}A, the $\eta_{ij}$ matrix is divided into blocks according to local communities. The diagonal blocks represent within-community interactions as in the case of a single community described above. Among those, the diagonal entries of the diagonal blocks (green) represent intraspecific effects that determine the persistence of a species over the next time in a given patch.  If the entries corresponding to the same species are identical among the patches, there is no environmental heterogeneity; otherwise, the differences among patches would describe the effects of such heterogeneity on that species.   As in the case of a single community, the non-diagonal entries of the diagonal blocks (blue) represent interspecific interactions, which can also vary among the patches. Finally, the diagonal entries of the non-diagonal blocks (red) represent possible dispersal of species between communities. The non-diagonal entries of the non-diagonal blocks (gray) are expected to be zero (otherwise, they would represent dispersal that changes species identities). On the other hand, the $\lambda_i$ parameters form a vector with multiple sections corresponding to different patches. These parameters represent the baseline colonization rates into the patches from the external species pool. If there are no differences among patches in these external inputs, then the entries of the vector $\lambda$ that correspond to the same species at different patches would be equal. Otherwise, differences among these entries could represent environmental variation, such as variation in abiotic conditions among patches that affect colonization success, independent of species interactions.

\subsection*{Examples of community structures}

\begin{figure*}[t]
\centering
\includegraphics[width=\textwidth]{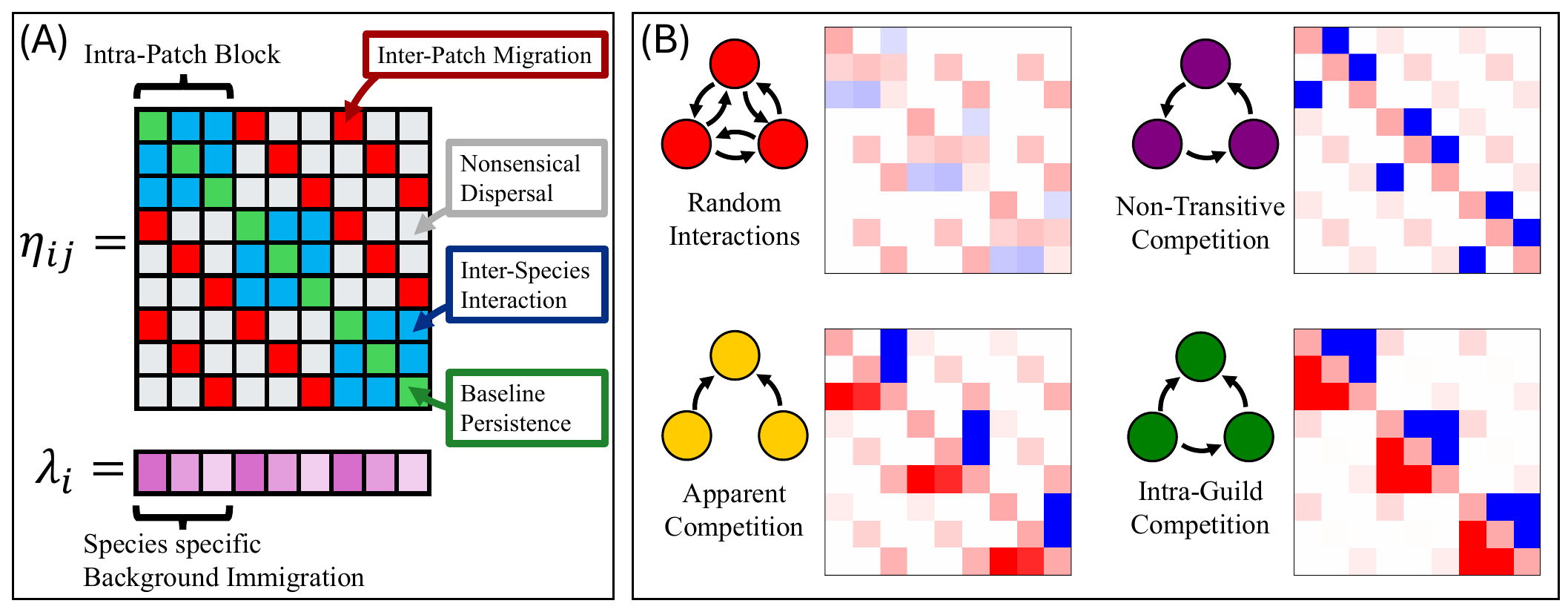}
\caption{\small Pattern of Maxiumum Caliber model parameters, interpretable as rates of ecological processes. (A) The $\eta_{ij}$'s represent the interactions between (blue) and within (green) species in the same patch, as well as rates of colonization among patches (red). The $\lambda_i$'s correspond to the rate of background colonization of a species into a patch. (B) Four different community structures (motifs) used to assess the potential utility of Maximum Caliber modeling. Each example system consists of three species found in three possible patches. The heatmaps show example $\eta_{ij}$ values (red for positive and blue for negative) for each motif.}
\label{fig:motifs}
\end{figure*}

To demonstrate the effectiveness of the Maximum Caliber approach, we consider four different community structures (community modules or motifs) for three species in each of three patches, as illustrated in Fig.~\ref{fig:motifs}B. The first example is a ``random interactions'' model inspired by random Lotka-Volterra models \citep{may:1972, allesina:2012, stone:2020}, where the within-community interspecific interactions are randomly chosen from a normal distribution with a given mean and variance. The second example is a trophic structure known as ``apparent competition'' \citep{holt:1977, holt:2017}, where two prey species share the same predator. The third example is known as ``intraguild predation'' \citep{holt:1997}, which includes a generalist consumer that can consume both a prey and its basal resource. In our last example, ``non-transitive competition'', each species is strongly excluded by one other species, as in a rock-paper-scissors game \citep{may:1975,sinervo:1996}. Fig.~\ref{fig:motifs}B shows the structure of the $\eta_{ij}$ matrix for each of these four types of metacommunity with three patches and randomly chosen dispersal parameters between them. (A graph representation of state transitions in each type of community is shown in Supp. Fig. S1.)

These examples superficially appear to have similar community structures, as one can be transformed into another by adding or flipping one arrow in their motif diagrams (Fig.~\ref{fig:motifs}B). However, they reveal qualitatively different dynamics in terms of how far they are from equilibrium (as will be quantified later). We use Maximum Caliber to model these case studies and to assess the ability of Maximum Caliber to correctly infer the parameters of the model.

\section*{Results}

\subsection*{Accuracy of parameter inference}

Maximum Caliber is particularly useful because it provides a method for inferring the parameters of the occupancy models, including identifying interaction structures (such as the examples in Fig.~\ref{fig:motifs}B).  An important concern is the degree to which such inferences are precise and unbiased.  We tested the inference of parameters using logistic regression from simulation data over a broad array of simulations that varied the number of patches, the number of species, and the duration $T$ of the time series. To do so, we simulated data resulting from selecting the parameters $\eta_{ij}$ and $\lambda_i$, subject to the assumptions of each motif, and then we used logistic regression on the simulation data to estimate the parameter values $\hat{\eta}_{ij}$ and $\hat{\lambda}_i$.  We next compared these estimated parameters with the original values $\eta_{ij}$ and $\lambda_i$ used to generate the simulations. We tested the accuracy of parameter inference using the random interactions model (the choice of parameters is described in the appendix); the other motifs yielded similar results.

\begin{figure*}[t]
\centering
\includegraphics[width=\textwidth]{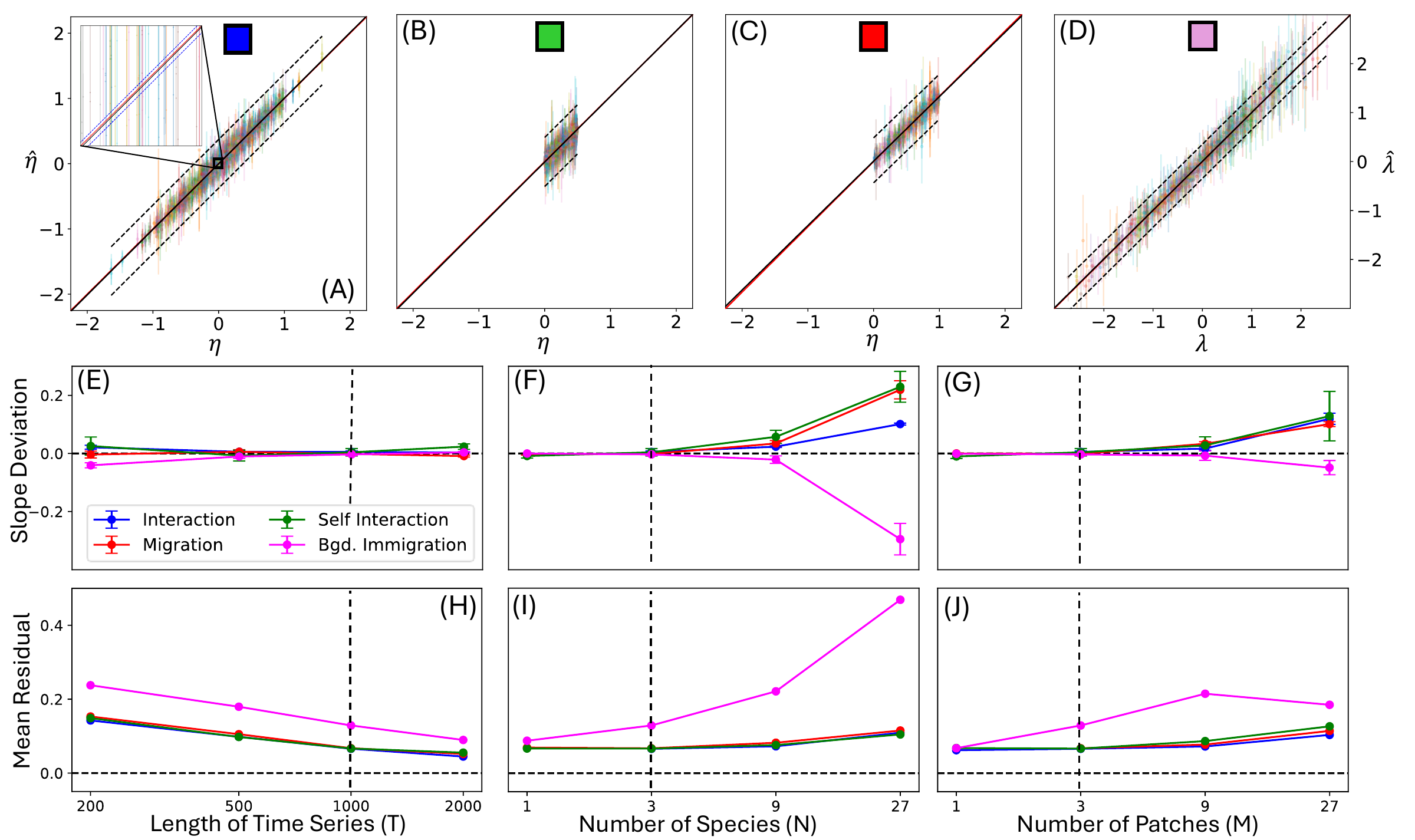}
\caption{\small Accuracy of parameter inference using logistic regression. (A)-(D) are plots of the inferred versus true parameter values for a system with the number of species $N=3$, the number of patches $M=3$, and the length of time series $T=1000$. The choice of random parameter values are described in the Appendix. The red line is slope 1, representing perfect inference. The black solid line is the line of fit (which largely overlaps with the red line). The dotted curves show the observational confidence interval for one standard deviation. Panel (A) shows the interspecific interactions, panel (B) shows the baseline persistence, panel (C) shows the migration rates, and panel (D) shows the background immigration rates (see Fig.~\ref{fig:motifs}A). The inset in panel (A) shows the mean confidence interval for the linear weighted regression (which is almost invisible in the main plot). Panels (E)-(J) show how the inference accuracy scales with the time period $T$ and system size $N$ and $M$. Panels (E), (F), and (G) show the error in the slope of weighted least squares fit of inferred parameter values against true values. Error bars are the standard error of the slope estimation. Panels (H), (I), and (J) show the mean difference between the inferred parameter values and their true values. Vertical dashed lines in (E)-(J) correspond to the parameters used in panels (A)-(D).}
\label{fig:inf_exam}
\end{figure*}

To quantify the accuracy of our inferences, we performed linear regression of the estimated values of the parameters on the original values. The slope and intercept of the regression  indicates whether there is bias in the parameter inference, and the residual variance measures the accuracy of such parameter estimates. Examples of such regression are shown in Fig.~\ref{fig:inf_exam}A--D. The slopes and intercepts are very close to 1 and 0, respectively, and the residual variances are small.  Crucially, these inferences appear equally good for all the parameters and indicate no bias in the estimates.

To test the scalability of our method, we simulated larger systems and inferred model parameters from different lengths of time series. We expected the accuracy of such inferences to depend positively on the length of the time series and negatively on system size (the number of interconnected patches and the number of species). Fig.~\ref{fig:inf_exam}E--J show how the slope and residual variance change with the length of the time series ($T$), number of species ($N$), and number of patches ($M$). These figures confirm our expectations and show that our parameter estimations are still accurate for many species and patches. Therefore, our method will be effective and tractable in applications to real systems.

\subsection*{Irreversibility and entropy production}

Maximum Caliber can model both steady-state dynamics that are at (or near) equilibrium and those that are not. To capture how far the dynamics is from equilibrium, we borrow the concept of ``entropy production'' from statistical physics, which can be defined as the relative entropy between the forward trajectories and the time-reversed trajectories \citep{peliti:2021}. This quantity has been used to measure the ``irreversibility'' of dynamical systems in physics and biology \citep{li:2024, grandpre:2024}. For a stationary process, the average rate of entropy production can be calculated as:
\begin{align}
    E_\text{p} = \sum_{\mathbf{x}, \mathbf{x}'} P(\mathbf{x}) P(\mathbf{x}' | \mathbf{x}) \log \frac{P(\mathbf{x}' | \mathbf{x})}{P(\mathbf{x} | \mathbf{x}')}
\end{align}
where $P(\mathbf{x})$ is the marginal probability of the system being in the state $\mathbf{x}$, and $P(\mathbf{x}' | \mathbf{x})$ is the conditional probability of the system transitioning from $\mathbf{x}$ to $\mathbf{x}'$ \citep{tome:2012}. The entropy production rate $E_\text{p}$ is zero when the system is at equilibrium, and is positive when the system is out of equilibrium. It is a general measure of non-equilibrium dynamics, independent of the Maximum Caliber framework.

Because the calculation of $E_\text{p}$ involves the ratio of transition probabilities, if these probabilities were to be estimated directly from observed time series, it would take a prohibitive amount of data. However, in our model, the entropy production can be calculated from the estimated values of the $\eta_{ij}$ and $\lambda_i$ parameters by using them to calculate the transition probabilities according to Eq.~[\ref{eq:cond-prob-1}]. This means that once a set of parameters has been inferred for a system, the entropy production can be calculated even if not all possible state transitions have been observed in a finite dataset, as long as the parameter inference is accurate.

Examples of the distribution of the entropy production for different realizations of each type of system are shown in Fig.~\ref{fig:ent_r2}A. We can see that the Maximum Caliber framework can capture a wide range of qualitatively different dynamics, from the near-equilibrium of the random interactions model, to the near-deterministic dynamics of the non-transitive competition model, and intermediates in between. The relative degrees of non-equilibrium in these motifs can be intuitively understood by comparing the amount of asymmetry in the interaction parameters $\eta_{ij}$. Indeed, Eq.~[\ref{eq:cond-prob-1}] is equivalent to the kinetic Ising model in statistical physics \citep{aguilera:2021}, and it is known that when the interactions are symmetric, the dynamics will reach an equilibrium with a time-reversible steady state. Among the motifs we study, the random interactions model is closest to equilibrium because the interactions are statistically symmetric, whereas the non-transitive competition model is farthest from equilibrium because the interactions are chosen to have large asymmetric values.

\subsection*{Predictability and Pseudo-$R^2$}

There is a distinction between how statistically significant the estimated parameters of a model are, and how accurately the model can predict future outcomes of a system. In linear regression, for example, we can have an estimated value of the slope that is significantly nonzero if we have a large amount of data. However, we may still have a large variance of the residual, which means that there is considerable uncertainty in our prediction even though our model parameters are accurate. One way to measure the predictability of a linear regression model is to use the coefficient of determination $R^2$, which is bounded between 0 and 1. The larger $R^2$ is, the greater the proportion of variation captured by the model, and the better the response variable is predicted by the explanatory variables.

Our Maximum Caliber model is equivalent to logistic regression; rather than continuous variables, we have categorical variables for both response and explanatory variables. To measure the variation captured by our model, we use a generalization of $R^2$ for logistic regression, called ``pseudo-$R^2$'' \citep{mittlbock:1996}. Several formulations of this quantity exist; a convenient one due to Cox \& Snell \citep{cox:1989} is:
\begin{align}
    R^2_\text{p} = 1 - \left( \frac{L_M}{L_0} \right)^{-2/n} = 1 - \mathrm{e}^{-D/n}
\end{align}
where $L_M$ is the likelihood of the model inferred by logistic regression (using maximum likelihood estimation), and $L_0$ is the likelihood of a null model that uses only the marginal probabilities of the response variables. Here $n$ is the number of data points, and $D = 2 \log (L_M / L_0)$ is known as the ``deviance'' of the inferred model. This definition of pseudo-$R^2$ is consistent with the standard $R^2$ in the case of linear regression. However, a major difference between the standard $R^2$ and pseudo-$R^2$ is that pseudo-$R^2$ ranges between 0 and a maximum value that is less than 1. For logistic regression, the maximum value is given by $R^2_\text{p,max} = 1 - L_0^{2/n} \leq 0.75$ \citep{nagelkerke:1991}. A modified definition of pseudo-$R^2$, due to Nagelkerke \citep{nagelkerke:1991}, normalizes $R^2_\text{p}$ by $R^2_\text{p,max}$, so that it ranges between 0 and 1; we will call this the ``normalized pseudo-$R^2$'', or $R^2_\text{np}$.

\begin{figure*}[t]
\centering
\includegraphics[width=.85\textwidth]{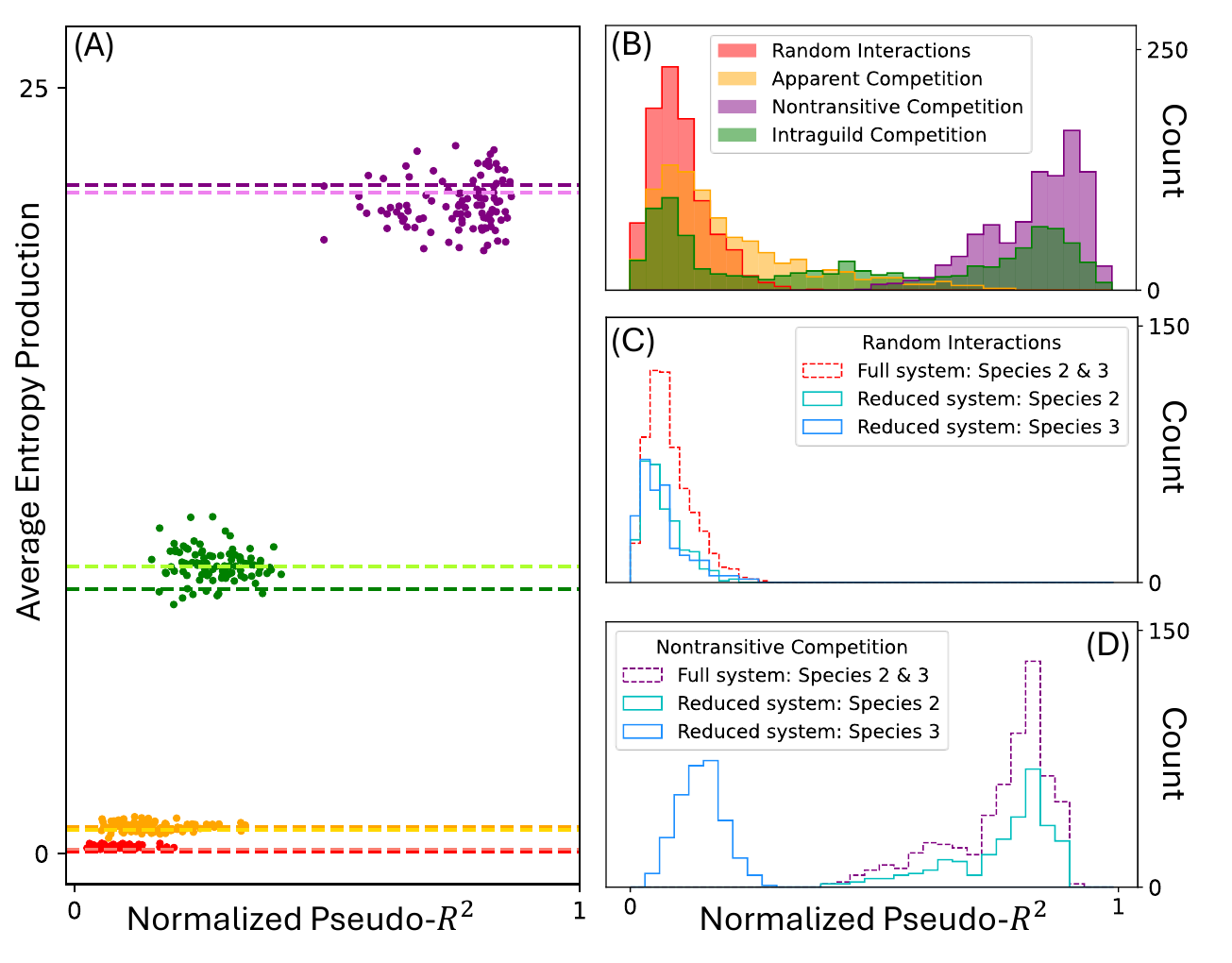}
\caption{\small Statistical measures of irreversibility and predictability of system dynamics. (A) The average rate of entropy production versus normalized pseudo-$R^2$. Colors represent different motifs in Fig.~\ref{fig:motifs}B; points represent replicate simulations with randomized parameter values. The larger the entropy production is, the further from equilibrium is a system. The larger the normalized pseudo-$R^2$ is, the bigger fraction of variation is captured by the Maximum Caliber model. (B) The distribution of $R_\text{np}^2$ values for each species-patch pair in systems with different community structures. (C)-(D) Comparison of the distribution of $R_\text{np}^2$ values between the full system and a reduced system that is missing one of the species (Species 1). Panel (C) shows results for the random interactions motif. Panel (D) shows the non-transitive competition motif, where} Species 3 is the one strongly inhibited by Species 1, and its predictability is strongly affected if Species 1 is missing from the data.
\label{fig:ent_r2}
\end{figure*}

We will use the normalized pseudo-$R^2$ to measure how well our inferred model makes predictions about the presence or absence of a species compared to using only its marginal probability. In Fig.~\ref{fig:ent_r2}B we show the distribution of the $R^2_\text{np}$ values for each of the four types of our simple community interaction structures. Note that the quantity $R^2_\text{p,max}$ measures how much variation there is in the data beyond the marginal probabilities, whereas $R^2_\text{np}$ measures how much of that variation is captured by our Maximum Caliber model (for their comparison, see Supp. Fig. S2). In general, we found that our Maximum Caliber model can capture a larger fraction of the variation in systems that are far from equilibrium than it can in systems that are near equilibrium. 

Moreover, the pseudo-$R^2$ value can be calculated for each species alone by using the likelihood function in Eq.~[\ref{eq:cond-prob-1}] for a single species. In Fig.~\ref{fig:ent_r2}B we show the distributions of $R^2_\text{np}$ for each species in different community structures. In the intraguild predation motif, for example, we see that not all species within a system are equally predictable. The top predator (the leftmost peak) turns out to have the lowest $R^2_\text{np}$ values and so is less predictable than the other two species. Variation in the $R^2_\text{np}$ values observed for any particular system comes from a combination of system structure, true parameter differences, and data realization (see Supp. Fig. S2).

\subsection*{Missing species}

In empirical studies, there is no guarantee that all species involved in ecological interactions are sampled.  Some species may be cryptic, or hard to identify, or may belong to taxonomic groups not sampled for pragmatic reasons.  Such missing data may affect predictions about the species that are actually observed. We can investigate the effect of such missing species using the simulations in the 3-species, 3-patch metacommunities explored above. To do so, we removed one species from the simulated data, then re-estimated the parameters of a model involving only the remaining two species. We examined the random interactions motif and the non-transitive competition motif, as they represent two extremes of community dynamics. Because there is symmetry among all three species within either motif, we removed the first species without loss of generality. To compare the results with the full model studied above, we re-calculated $R_\text{np}^2$ values for each of the two remaining species. Shown in Fig.~\ref{fig:ent_r2}C,D is a comparison of the original distribution of $R_\text{np}^2$ values for these two species in the full model and the recalculated $R_\text{np}^2$ values after one species is removed. In the case of random interactions (Fig.~\ref{fig:ent_r2}C), we see only a slight shift of $R_\text{np}^2$ to lower values for the remaining species.  Thus, missing any particular species in a system of this type only weakly affects predictions for the remaining species. With non-transitive competition (Fig.~\ref{fig:ent_r2}D), we see that missing species 1 in the data has almost no effect on the predictions made about species 2, but a strong effect on species 3. This is because species 1 strongly excludes species 3, making it an important factor in determining the presence of species 3, but it has a weaker impact on species 2. Our results suggest that, when species are missing or not included in empirical data, effects on the predictability of the other species are not uniform.

\section*{Discussion}

We have introduced the method of Maximum Caliber for community ecology, as illustrated by a metacommunity occupancy model that corresponds to previously well-studied scenarios.  There are two important advances made possible by Maximum Caliber.  The most important of these is that it allows for unbiased inferences of dynamic parameters from observed spatio-temporal data using simple logistic regression.  A second important advantage is that inference is not limited to steady states that are at equilibrium, but that it can be carried out even for non-equilibrium dynamics.  As the relevance of steady-state and equilibrium assumptions in ecology are increasingly questioned \citep{mccann:2000, chesson:2024, morozov:2024, spencer:2025}, our method may provide a useful approach to study community ecology, such as in applications where human impacts are rapidly increasing.  The use of simple logistic regression also means that the resulting parameters are easily interpretable in terms of biological processes driving the dynamics of (meta)communities.

We have also introduced two useful metrics to characterize metacommunity dynamics.  One is the entropy production rate of the system, calculated from observed data.  This helps quantify how far-from-equilibrium (or irreversible) a given system might be, and helps identify systems that may undergo assembly cycles (such as in non-transitive competition).  The other useful metric is pseudo-$R^2$ which can be calculated for the system as a whole and for each species separately.  This metric quantifies the predictability of future states, given the inferred dynamics of the system.  Because we can calculate this quantity for each species, it provides a tool that can reveal the ``internal structure'' of the metacommunity (\textit{sensu} \citep{leibold:2022}).  In conservation ecology, this can help identify species that are particularly robust (or sensitive) to perturbations, thereby informing the crafting of conservation measures \citep{horvath:2025}. The metric may also be be more broadly useful in the increasingly urgent issue of ecological forecasting \citep{petchey:2006}.

\subsection*{Potential issues}

When applying the method of Maximum Caliber, some practical issues need to be considered. First is the amount of available data. Inferring the \emph{dynamics} of metacommunities requires long time series. As shown by our simulation results, at least tens or hundreds of time points are needed to infer parameters with any confidence. This number increases with the size of the species pool and the number of patches in the metacommunity. One way to gather a sufficient number of data points is to have replicate experiments or replicate time series, even if each time series does not run for a long time. The parameter inference method is such that one can compile data points from different trajectories together in a single logistic regression, as long as it is reasonable to assume that the interaction parameters are the same across these trajectories.

Even for short time series from which parameter estimation becomes less accurate, our method can still help detect general trends in the ecological processes. For example, we may correlate the inferred parameter values with known environmental variables. A simulated example is shown in Fig.~S3 with parameters inferred from short time series of realistic lengths. Although the estimated value of each parameter is noisy, it is nevertheless possible to detect a correlation between the inferred parameters and the simulated environmental variables. This shows that our method can be useful for extracting ecological insight from even short time series data. We note that even short time series would potentially contain more information than comparable static data. In addition to the static marginal probabilities, our method uses the temporal order of data points, which makes the results more informative than atemporal traditional methods.

Another issue is missing species. The causal role of missing species might be folded into  ``effective parameters'' between the detected species, absorbing some of the effects of the missing species. We have seen above that, even with missing species, the dynamics of some of the monitored species can still be predicted quite accurately, as measured by the normalized pseudo-$R^2$ metric for these species. In our examples, the predictability of a species is significantly reduced if the missing species exerts strong interactions on that species. The generality of this trend needs further study.  A related issue is when a species is always (or almost always) present or absent everywhere in the community. This would cause the regression algorithm to falter because of the lack of variation in the input value. In this case, it would be useful to remove that species from the pool. The effect of that species will then be absorbed into the background (i.e., the $\lambda_i$ parameters of the other species). In practice, the species pool can be pre-screened to remove such ``passive'' species that act as a background for the other species. The problem of false absences in the data due to undetected species is not specific to our method. Since this problem mainly involves species that are locally rare (though presumably common elsewhere in the metacommunity), we expect that such false absences may not strongly affect the transitions of other species (i.e., the $\lambda_i$ and $\eta_{ij}$ parameters of the other species).

\subsection*{Generalizing the model}

When using the Maximum Caliber approach to craft metacommunity models, we have chosen to use the species occupancy frequency $f_i$ and the cross-correlation $c_{ij}$ at consecutive time points as constraints. These are convenient choices, in that they are readily calculated from temporal presence-absence data, and in that the resulting model has a simple parametric form and ecological interpretation. A similar presence-absence model for community assembly has been introduced and analyzed in \citep{fisher:2014, dickens:2016}. However, the choices for constraints are not unique, and different choices could lead to different parameterized models. For instance, one could calculate the cross-correlation between two species at longer time lags than consecutive time points. Including such a new constraint in the maximization of trajectory entropy will lead to a new term in the logistic function (Eq.~\ref{eq:cond-prob-1}) that may represent species interactions over longer timescales. In practice, one could try including time-lagged variables in the logistic regression and test whether they are significant, similar to testing for Granger causality \citep{tank:2021}. This will likely require a more substantial amount of data.

Similarly, because we used cross-correlation between \emph{pairs} of species as constraints, the resulting Maximum Caliber model included only pairwise interactions between different species. Higher-order interactions, such as those induced by trait-mediated indirect effects \citep{ohgushi:2012,werner:2003}, can involve three or more species, with important consequences in some ecosystems \citep{mickalide:2019,goudard:2008}. Such interactions can be incorporated into our approach by imposing higher-order statistics (higher mixed moments) as constraints in building a Maximum Caliber model. It is possible to have many forms of such higher-order interactions, even among three species, by considering different combinations of present and time-lagged variables, such as $x_i^{(t+2)} x_j^{(t+1)} x_k^{(t)}$, $x_i^{(t+1)} x_j^{(t+1)} x_k^{(t)}$, or $x_i^{(t+1)} x_j^{(t)} x_k^{(t)}$. Models with such terms will become increasingly complex and require an extensive amount of data to make reliable inferences; their interpretation in terms of ecological processes will also be more challenging. Yet, there are methods from Maximum Entropy models that can potentially be used to selectively introduce such terms based on their statistical significance \citep{cayco-gajic:2018}.

\subsection*{Future work}

There are many areas for future work using Maximum Caliber that one can imagine. It could for instance be used to study ecological succession, such as how secondary succession compares to primary succession, how multiple trophic levels and their interactions affect community assembly, and how the assembly of microbes in host-associated microbiomes is determined.  Our method could be applied to any number of existing longitudinal community studies such as those from Long Term Ecological Research (LTER) programs and National Ecological Observatory Networks (NEON); novel monitoring studies promise that such data will be increasingly available in the future. One such arena where, in the future, long time-series might in particular be abundantly available is in microbiome research, where ``patches'' are individual hosts harboring communities of symbionts. 

An important consideration in preparing the data for analysis is sampling frequency. Although our method can be used to model dynamical processes, it could miss features if sampling is too sparse, leaving out dynamical details in between measured time points. Conversely, sampling too frequently and modeling interactions at only consecutive time points may miss out processes happening over longer time scales. Some systems may have clear intrinsic timescales, such as those with seasonal dynamics, in which case the choice of sampling frequency is more obvious.

In summary, Maximum Caliber provides a novel, versatile framework for the analysis of metacommunity dynamics involving diverse biotas interacting in complex landscapes that feature environmental heterogeneity, spatial organization, and stochasticity.  It provides both dynamical models that can make detailed predictions about patterns, and a tool for inferring parameters from a common type of metacommunity data.  In doing so, Maximum Caliber promises to substantially move the field of community ecology in ways that can provide better insights about nature and facilitate its wise management and conservation.

\section*{Appendix}

\subsection*{Derivation of the Maximum Caliber solution}

To maximize the function in Eq.~(\ref{eq:Lagrange}), we set its derivative with respect to $P(\Gamma)$ equal to zero:
\begin{align}
\frac{\partial \mathcal{L}}{\partial P(\Gamma)} = &- \log P(\Gamma) - 1 + \sum_{ij} \tilde{\eta}_{ij} \, \frac{1}{T} \sum_t x_i^{(t+1)} x_j^{(t)}\nonumber\\
&+ \sum_i \tilde{\lambda}_i \, \frac{1}{T} \sum_t x_i^{(t)} + \mu = 0 .
\end{align}
Solving for $P(\Gamma)$ gives
\begin{align} \label{eq:P_Gamma}
P(\Gamma) = \exp \left( \frac{1}{T} \sum_t \left( \sum_{ij} \tilde{\eta}_{ij} \, x_i^{(t+1)} x_j^{(t)} + \sum_i \tilde{\lambda}_i \, x^{(t)}_i \right) + \mu - 1 \right) .
\end{align}
The parameter $\mu$ can be fixed by the normalization of probability, yielding:
\begin{equation}
\mu = 1 - \log \sum_{\Gamma} \exp \left( \frac{1}{T} \sum_t \left( \sum_{ij} \tilde{\eta}_{ij} \, x_i^{(t+1)} x_j^{(t)} + \sum_i \tilde{\lambda}_i \, x^{(t)}_i \right) \right) .
\end{equation}
Recall that $P(\Gamma)$ is the probability of a trajectory, $\Gamma = \{ \mathbf{x}^{(0)} \cdots \mathbf{x}^{(T)} \}$. The form of Eq.~(\ref{eq:P_Gamma}) implies that the process is Markovian \citep{ge:2012}, i.e., $P \big( \mathbf{x}^{(t+1)} \big| \mathbf{x}^{(0)} \cdots \mathbf{x}^{(t)} \big) = P \big( \mathbf{x}^{(t+1)} \big| \mathbf{x}^{(t)} \big)$ (see detailed proof in Supplementary Material). Therefore, we can rewrite the probability of a trajectory as a product of conditional probabilities: 
\begin{equation}
P(\Gamma) \equiv P \big( \mathbf{x}^{(0)} \cdots \mathbf{x}^{(T)} \big) = P \big( \mathbf{x}^{(0)} \big) \prod_{t=0}^{T-1} P \big( \mathbf{x}^{(t+1)} \big| \mathbf{x}^{(t)} \big) .
\end{equation}
It can be shown that the conditional probability for each time step is given by: 
\begin{align} \label{eq:cond-prob}
P \big( \mathbf{x}^{(t+1)} \big| \mathbf{x}^{(t)} \big) = \frac{1}{Z_t} \exp \left( \sum_{ij} \eta_{ij} \, x_i^{(t+1)} x_j^{(t)} + \sum_i \lambda_i \, x^{(t+1)}_i \right) ,
\end{align}
where $\eta_{ij}$ and $\lambda_i$ are parameters defined in terms of the original parameters $\tilde{\eta}_{ij}$ and $\tilde{\lambda}_i$ (see Supplementary Material), and $Z_t$ is a normalization factor. We can further write Eq.~(\ref{eq:cond-prob}) as a product of conditional probabilities for each species, i.e., $P \big( \mathbf{x}^{(t+1)} \big| \mathbf{x}^{(t)} \big) = \prod_i P \big( x_i^{(t+1)} \big| \mathbf{x}^{(t)} \big)$ with
\begin{equation}
P \big( x_i^{(t+1)} \big| \mathbf{x}^{(t)} \big) = \frac{1}{Z_{t,i}} \exp \left( x^{(t+1)}_i  \left( \sum_{j} \eta_{ij} \, x_j^{(t)} + \lambda_i \right) \right) .
\end{equation}
Fixing the normalization factor $Z_{t,i}$ leads to the result in Eq.~[\ref{eq:cond-prob-1}].

\subsection*{Choice of parameter values}

In the production of data used to produce Fig.\ref{fig:inf_exam}, the $\eta_{ij}$ on the main diagonal (green in Fig. \ref{fig:motifs}A) are taken from a uniform distribution between 0 and 0.5. The diagonal of each diagonal block is made identical so that the baseline persistence parameters are only species specific. This ensures that a species is more likely to be present at the next time step, if it is present in the current time step. The off-diagonal entries of the diagonal blocks (blue in Fig. \ref{fig:motifs}A) are taken from a normal distribution with mean 0 and a standard deviation of 0.5. To assign the values of $\eta_{ij}$ that are associated with migration (red in Fig. \ref{fig:motifs}A), first, each pair of patches is connected with a probability of $1/M$. Then, for each pair of connected patches, the diagonal of the block associated with that pair is filled with parameters chosen from a uniform distribution between 0 and 1. The $\lambda_i$ for each species are drawn from a normal distribution with mean 0 and standard deviation of 1. To include environmental variation between patches, an additional random value is added to all $\lambda_i$ in each patch, taken from a normal distribution centered at 0 with a standard deviation of 0.2. This ensures that the same species in different patches have similar but not identical interactions with the abiotic factors in their environment.

In the production of data used to produce Fig. \ref{fig:ent_r2}, the parameters are chosen in accordance with each motif. Each system has 3 species and 3 patches, and the patches are all connected. For the random interactions motif (red in Fig. \ref{fig:motifs}B), the parameters are chosen in the same manner as above except without the environmental variation added to the $\lambda_i$. In the other motifs, the diagonal $\eta_{ij}$ and the $\lambda_i$ are set to 1. In the apparent competition motif (yellow in Fig. \ref{fig:motifs}B), for interspecific interaction parameters (blue in Fig. \ref{fig:motifs}A), a random value $s$ is chosen, and then in each diagonal block we set $\eta_{02} = \eta_{12} = -s$, $\eta_{20} = s$, and $\eta_{21} = s/2$. For the intraguild predation motif (green in Fig. \ref{fig:motifs}B), three values $a$, $b$, and $c$ are chosen, and then in each diagonal block we set $\eta_{20} = -\eta_{02} = a$, $\eta_{21} = -\eta_{12} = b$, and $\eta_{10} = -\eta_{01} = c$. For the non-transitive competition motif (purple in Fig. \ref{fig:motifs}B), a value $s$ is chosen, and then in each diagonal block we set $\eta_{01} = \eta_{12} = \eta_{20} = -s$. In the case of Fig. \ref{fig:ent_r2}(A), for instance, $s=10$.

%\bibliographystyle{plainnat}
%\bibliography{max_cal_ref_1}

\clearpage
\appendix

\setcounter{equation}{0}
\renewcommand\theequation{S\arabic{equation}}

\setcounter{figure}{0}
\renewcommand\thefigure{S\arabic{figure}}

\section*{Supplementary Material}

\subsection*{Detailed derivation of the Maximum Caliber model}

For simplicity and clarity, we will derive the Maximum Caliber solution for a single species, which means dropping the subscript $i$. We start from the loss function:
\begin{align}
L=&-\sum_{\{x^{(t)}\}}P(\{x^{(t)}\}) \log P({\{x^{(t)}\}}) + \eta\left(\sum_{\{x^{(t)}\}}P(\{x^{(t)}\})\frac{1}{T}\sum_{t=1}^{T-1}x^{(t+1)}x^{(t)}-c\right)\nonumber\\
&+\lambda \left(\sum_{\{x^{(t)}\}}P(\{x^{(t)}\}) \frac{1}{T}\sum_{t=1}^{T-1}x^{(t)}-f\right)
+ \mu \left(\sum_{\{x^{(t)}\}}P(\{x^{(t)}\})-1\right)
\end{align}
maximizing with respect to $P(\{x^{(t)}\})$ gives
\begin{align}
\frac{\partial L}{\partial P(\{x^{(t)}\})}= -\log P(\{x^{(t)}\})-1 + \frac{\eta}{T} \sum_{t=1}^{T-1}x^{(t+1)}x^{(t)} + \frac{\lambda}{T}\sum_{t=1}^{T-1}x^{(t)} + \mu = 0
\end{align}
if we renormalize our parameters with $\eta \leftarrow \eta/T$, $\lambda \leftarrow \lambda/T$, and $Z \leftarrow e^{1-\mu}$ and solve for $P(\{x^{(t)}\})$,
\begin{align}
P(\{x^{(t)}\})= \frac{1}{Z} \exp\left(\eta\sum_{t=1}^{T-1}x^{(t+1)}x^{(t)} + \lambda\sum_{t=1}^{T-1}x^{(t)}\right) = \frac{1}{Z} \prod_{t=1}^{T-1}\exp \left(\eta x^{(t+1)}x^{(t)} + \lambda x^{(t)}\right)
\end{align}
To see why this is a Markovian process, we will calculate the conditional probability starting from the end of the time series and working backwards. The joint probability of the full trajectory from $t=1$ to $T$ is given by a product of terms for each time step:
\begin{align}
P\left(x^{(1)} \cdots x^{(T)}\right)=\frac{1}{Z}\exp\left(\eta x^{T}x^{T-1}+\lambda x^{T-1}\right)\exp\left(\eta x^{T-1}x^{T-2}+\lambda x^{T-2}\right)\cdots\exp\left(\eta x^1x^0+\lambda x^0\right)
\end{align}
We then calculate the joint probability of a partial trajectory from $t=1$ to $T-1$ by summing over the possible states of $x^{(T)}$
\begin{align}
P\left(x^{(1)} \cdots x^{(T-1)}\right)&=\sum_{x^{(T)}=0}^1 P\left(x^{(1)} \cdots x^{(T)}\right)\nonumber\\ &= \frac{1}{Z}\left(1+\exp\left(\eta x^{T-1}\right)\right) \exp\left(\lambda x^{T-1}\right)\exp\left(\eta x^{T-1}x^{T-2}+\lambda x^{T-2}\right)\cdots
\end{align}
We can use these to calculate the conditional probability of $x^{(T)}$ given the trajectory up to $T-1$.
\begin{align}
P\left(x^{(T)}|x^{(1)} \cdots x^{(T-1)}\right) = \frac{P\left(x^{(1)} \cdots x^{(T)}\right)}{P\left(x^{(1)} \cdots x^{(T-1)}\right)} =\frac{\exp(\eta x^{(T)}x^{(T-1)})}{1+\exp(\eta x^{(T-1)})}
\end{align}
From this expression we see that the conditional probability depends only on $x^{(T-1)}$, which means this step is Markovian. We can rewrite this conditional probability as $P\left(x^{(T)}|x^{(T-1)}\right)$ and, for $x^{(T)}=1$,:
\begin{equation}
P\left(x^{(T)} \!=\! 1|x^{(T-1)}\right) = \frac{\exp(\eta x^{(T-1)})}{1+\exp(\eta x^{(T-1)})} = \frac{1}{1+\exp(-\eta x^{(T-1)})}
\end{equation}
Through a similar process, we can calculate the joint probability of the trajectory from $t=1$ to $T-2$:
\begin{equation}
P\left(x^{(T)}|x^{(1)} \cdots x^{(T-1)}\right) = ...
\end{equation}
Then we calculate the conditional probability of $x^{(T-1)}$ given the trajectory up to $t=T-2$:
\begin{align}
P\left(x^{(T-1)}|x^{(1)} \cdots x^{(T-2)}\right) &= \frac{P\left(x^{(1)} \cdots x^{(T-1)}\right)}{P\left(x^{(1)} \cdots x^{(T-2)}\right)}\nonumber\\ &=\frac{\left(1+\exp(\eta x^{(T-1)})\right)\exp(\lambda x^{(T-1)} +\eta x^{(T-1)}x^{(T-2)})}{2+(1+\exp(\eta))\exp(\lambda+\eta x^{(T-2)})}
\end{align}
Again we see that this conditional probability depends only on $x^{(T-2)}$, which means it is Markovian. This conditional probability can be simplified as
\begin{align}
P\left(x^{(T-1)} \!=\! 1|x^{(1)} \cdots x^{(T-2)}\right)= \frac{1}{1+\exp(-\eta x^{(T-2)}-\lambda')}
\end{align}
where $\lambda'= \lambda+\log(1+\exp(\eta)/2)$. By continuing this recursion we obtain
\begin{align}
P\left(x^{(T-2)}=1|x^{(T-3)}\right)= \frac{1}{1+\exp(-\eta x^{T-3}-\lambda'')}
\end{align}
with $\lambda''=\lambda+\log\left(\frac{1+\exp(\eta+\lambda')}{1+\exp(\lambda')}\right)$. As we move further away from the end of the time series, the effective parameter value of $\lambda$ quickly approaches a constant. To find this limiting value, $\tilde{\lambda}$, we can set $\lambda'=\lambda''$ in the recursive relation:
\begin{align}
\exp(\tilde{\lambda})=\exp(\lambda)\frac{1+\exp(\eta+\tilde{\lambda})}{1+\exp(\tilde{\lambda})}
\end{align}
Solving this for $\tilde{\lambda}$ gives
\begin{align}
\exp\left(\tilde{\lambda}\right)=\frac{1}{2}\exp(\eta+\lambda)-1+\sqrt{(1-\exp(\eta+\lambda))^2+4exp(\lambda)}\label{eq:lambdatil}
\end{align}
Therefore, for large $T$, the conditional probability at each time step can be taken as:
\begin{align}
P\left(x^{(t+1)}=1|x^{(t)}\right)= \frac{1}{1+\exp(-\eta x^{(t)}-\tilde{\lambda})}
\end{align}

Note that when calculating the average occupancy of a species $f_i$ to use in the Maximum Caliber calculation, there is a choice of whether to sum from $x^{(1)}$ to $x^{(T-1)}$, or from $x^{(2)}$ to $x^{(T)}$. More generally, we can use
\begin{align}
f_i=\frac{1}{T}\sum_{t=1}^{T-1} \alpha x_i^{(t)} + (1-\alpha) x_i^{(t+1)}
\end{align}
for any value of $\alpha$ between $0\leq\alpha\leq1$. Our choice above corresponds to $\alpha = 1$. Following a similar scheme for $\alpha=0$, we find that $\lambda$ converges to the same value. As such, for long time series (large $T$), the resulting model will be the same regardless of the value of $\alpha$ used.

\subsection*{Maximum Caliber versus Logistic Regression}

Here we give a comparison between the parameter inference in Maximum Caliber and in logistic regression. In Maximum Caliber, strictly speaking, we need to find parameter values by solving the constraint equations:
\begin{align}
f_{i}^\textrm{obs}&= \sum_{\{x_i^{(t)}\}} P(\{x_i^{(t)}\})\frac{1}{T} \sum_t x_i^{(t+1)}\\
&=\frac{1}{T} \sum_t \sum_{x_i^{(t+1)}} P(x_i^{(t+1)})x_i^{(t+1)}\\
&=\frac{1}{T} \sum_t \sum_{x_i^{(t+1)}} \sum_{\mathbf{x}^{(t)}} P(x_i^{(t+1)}|\mathbf{x}^{(t)}) P(\mathbf{x}^{(t)}) x_i^{(t+1)}
\end{align}
If we assume that the process is stationary, then each $t$ term would give the same result, which means:
\begin{align}
f_{i}^\textrm{obs}&= \sum_{x_i^{(t+1)}} \sum_{\mathbf{x}^{(t)}} P(x_i^{(t+1)}|\mathbf{x}^{(t)}) P(\mathbf{x}^{(t)})  x_i^{(t+1)}\\
&=\sum_{\mathbf{x}^{(t)}} P(\mathbf{x}^{(t)}) P(x_i^{(t+1)}\!=\!1|\mathbf{x}^{(t)})\label{eq:f_maxcal}
\end{align}
Similarly for $c_{ij}$:
\begin{align}
c_{ij}^\textrm{obs}&= \sum_{\{x_i^{(t)}\}} P(\{x_i^{(t)}\})\frac{1}{T} \sum_t x_i^{(t+1)}x_j^{(t)}\\
&=\frac{1}{T} \sum_t \sum_{x_i^{(t+1)}} \sum_{\mathbf{x}^{(t)}} P(x_i^{(t+1)},x_j^{(t)}) x_i^{(t+1)} x_j^{(t)}\\
&=\frac{1}{T} \sum_t \sum_{x_i^{(t+1)}} \sum_{\mathbf{x}^{(t)}} P(x_i^{(t+1)}|\mathbf{x}^{(t)}) P(\mathbf{x}^{(t)}) x_i^{(t+1)}x_j^{(t)}\\
&= \sum_{x_i^{(t+1)}} \sum_{\mathbf{x}^{(t)}} P(x_i^{(t+1)}|\mathbf{x}^{(t)}) P(\mathbf{x}^{(t)}) x_i^{(t+1)}x_j^{(t)}\\
&=\sum_{\mathbf{x}^{(t)}} P(x_i^{(t+1)}\!\!=\!\!1|\mathbf{x}^{(t)}) P(\mathbf{x}^{(t)})x_j^{(t)}\\
&=\sum_{\mathbf{x}^{(t)}} P(\mathbf{x}^{(t)}) P(x_i^{(t+1)}\!\!=\!\!1|\mathbf{x}^{(t)}) x_j^{(t)}\label{eq:c_maxcal}
\end{align}

In logistic regression, we minimize the cross entropy between the predicted probabilities $P_t$ and the realized probabilities $y_t$.
\begin{align}
L=-\sum_t y_t \log P_t +(1-y_t)\log(1-P_t)
\end{align}
where $y_t$ is the presence or absence of a species in a patch at the next time step $x_i^{(t+1)}$, and $P_t$ is the probability of that species being present in that patch given the configuration of the system at the previous time step. 
\begin{align}
P_t&=\prod_iP(x_i^{(t+1)}\!\!=\!\!1|\mathbf{x}^{(t)})=\prod_i\frac{1}{1+e^{-\sum_j\eta_{ij}x_j^{(t)}-\lambda_i}}
\end{align}
With this expression, the log loss becomes:
\begin{align}
L=-\sum_t\sum_i\left(x_i^{(t+1)}\log P(x_i^{(t+1)}|\mathbf{x}^{(t)})+(1-x_i^{(t+1)})\log (1-P(x_i^{(t+1)}|\mathbf{x}^{(t)}))\right)
\end{align}
Differentiating the loss with respect to $\lambda_i$ gives
\begin{align}
\frac{\partial L}{\partial \lambda_i} &= \sum_tx_i^{(t+1)}\left(1-P\left(x_i^{(t+1)}|\mathbf{x}^{(t)}\right)\right)-\left(1-x_i^{(t+1)}P\left(x_i^{(t+1)}|\mathbf{x}^{(t)}\right)\right)\\
&=-\sum_t x_t^{(t+1)}-P\left(x_i^{(t+1)}\!\!=\!\!1|\mathbf{x}^{(t)}\right)\\
&=-T\left(f_i^\textrm{obs}-\frac{1}{T} \sum_t P\left(x_i^{(t+1)}\!\!=\!\!1|\mathbf{x}^{(t)}\right)\right)
\end{align}
Differentiating with respect to $\eta_{ij}$ gives
\begin{align}
\frac{\partial L}{\partial \eta_{ij}}&=-\sum_t x_i^{(t+1)}x_j^{(t)}(1-P\left(x_i^{(t+1)}|\mathbf{x}^{(t)}\right))-\left(1-x_i^{(t+1)}\right)x_j^{(t)}P\left(x_i^{(t+1)}|\mathbf{x}^{(t)}\right)\\
&=-\sum_t \left(x_i^{(t+1)}-P\left(x_i^{(t+1)}\!\!=\!\!1|\mathbf{x}^{(t)}\right)\right)x_j^{(t)}\\
&=-T\left(c_{ij}^\textrm{obs}-\frac{1}{T}\sum_tP\left(x_i^{(t+1)}\!\!=\!\!1|\mathbf{x}^{(t)}\right)x_j^{(t)}\right)
\end{align}
After maximizing the log likelihood, we should get 
\begin{align}
f_i^\textrm{obs} &= \frac{1}{T} \sum_t P\left(x_i^{(t+1)}\!\!=\!\!1|\mathbf{x}^{(t)}\right)\label{eq:f_log}\\
c_{ij}^\textrm{obs} &= \frac{1}{T}\sum_tP\left(x_i^{(t+1)}\!\!=\!\!1|\mathbf{x}^{(t)}\right)x_j^{(t)}\label{eq:c_log}
\end{align}
Note that in these equations the $x_i^{(t)}$'s are the observed values of species presence or absence, rather than dummy indices that are summed over in Eqs.~(\ref{eq:f_maxcal}) \& (\ref{eq:c_maxcal}).

If we compare Eqs.~(\ref{eq:f_log}) \& (\ref{eq:c_log}) to Eqs.~(\ref{eq:f_maxcal}) \& (\ref{eq:c_maxcal}), the difference between the logistic regression calculation and the strict Maximum Caliber calculation is that logistic regression takes an average over time of observed data, whereas Maximum Caliber takes an average over the probability of system trajectories. When the time series are sufficiently long, the two calculations will give equal results.

\clearpage
\subsection*{Supplementary Figures}

\begin{figure}[h!]
\centering
\includegraphics[width=.95\textwidth]{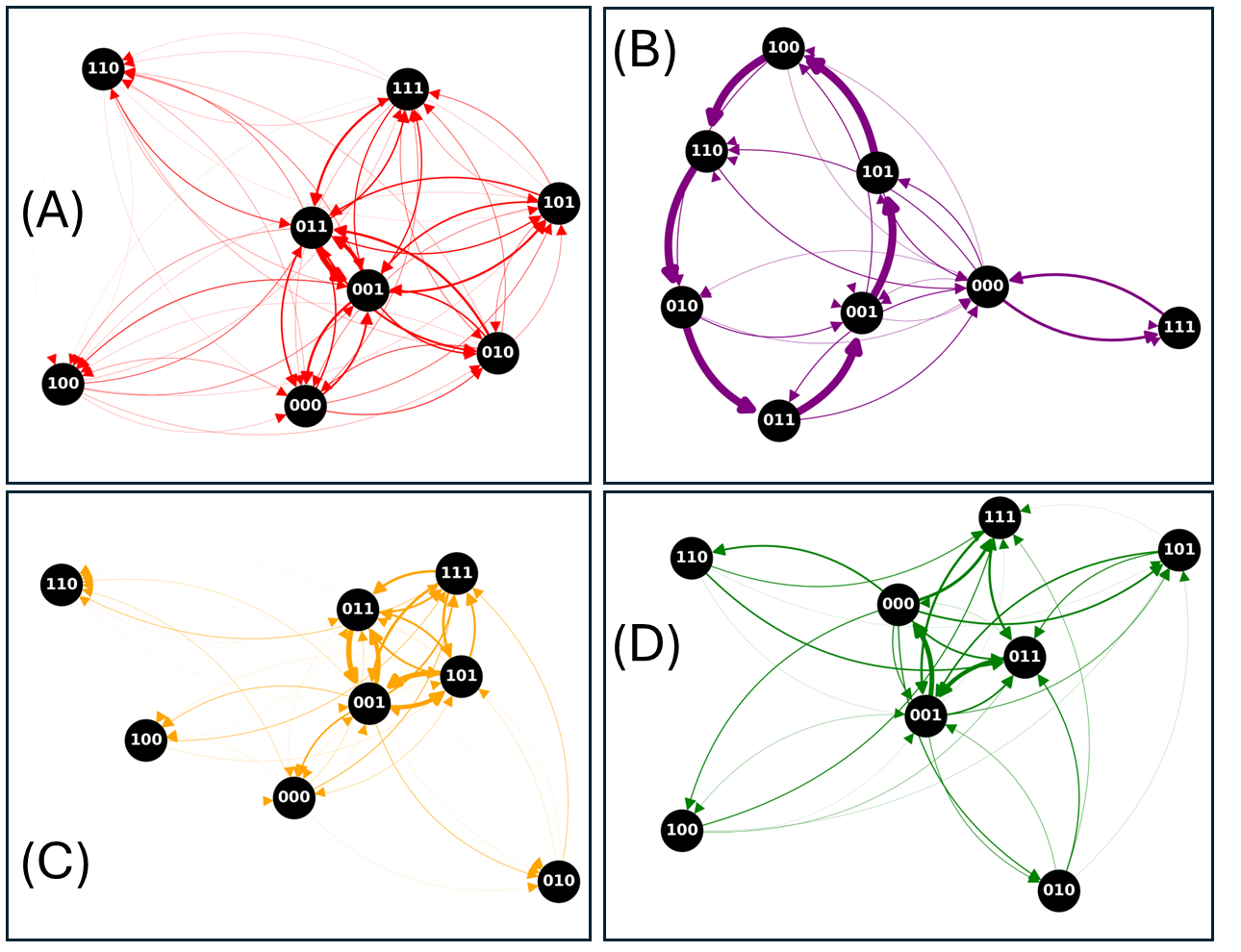}
\caption{Graph representations of the state transitions in the four motifs studied in the main text: (A) random interactions, (B) non-transitive competition, (C) apparent competition, and (D) intraguild competition. Each of the systems is a single patch with three species. Each node represents a state of the system labeled by a binary vector that denotes whether each species is present. The thickness of an arrow between two nodes corresponds to the rate of transition between the states, calculated from simulations of over 10000 time steps. The thickness is relatively homogeneous in the random interactions motif, showing that the system is near equilibrium. In contrast, there is a dominant cycle in the non-transitive competition motif, showing that the system is far from equilibrium.}
\label{fig:networks}
\end{figure}

\begin{figure}
\centering
\includegraphics[width=.95\textwidth]{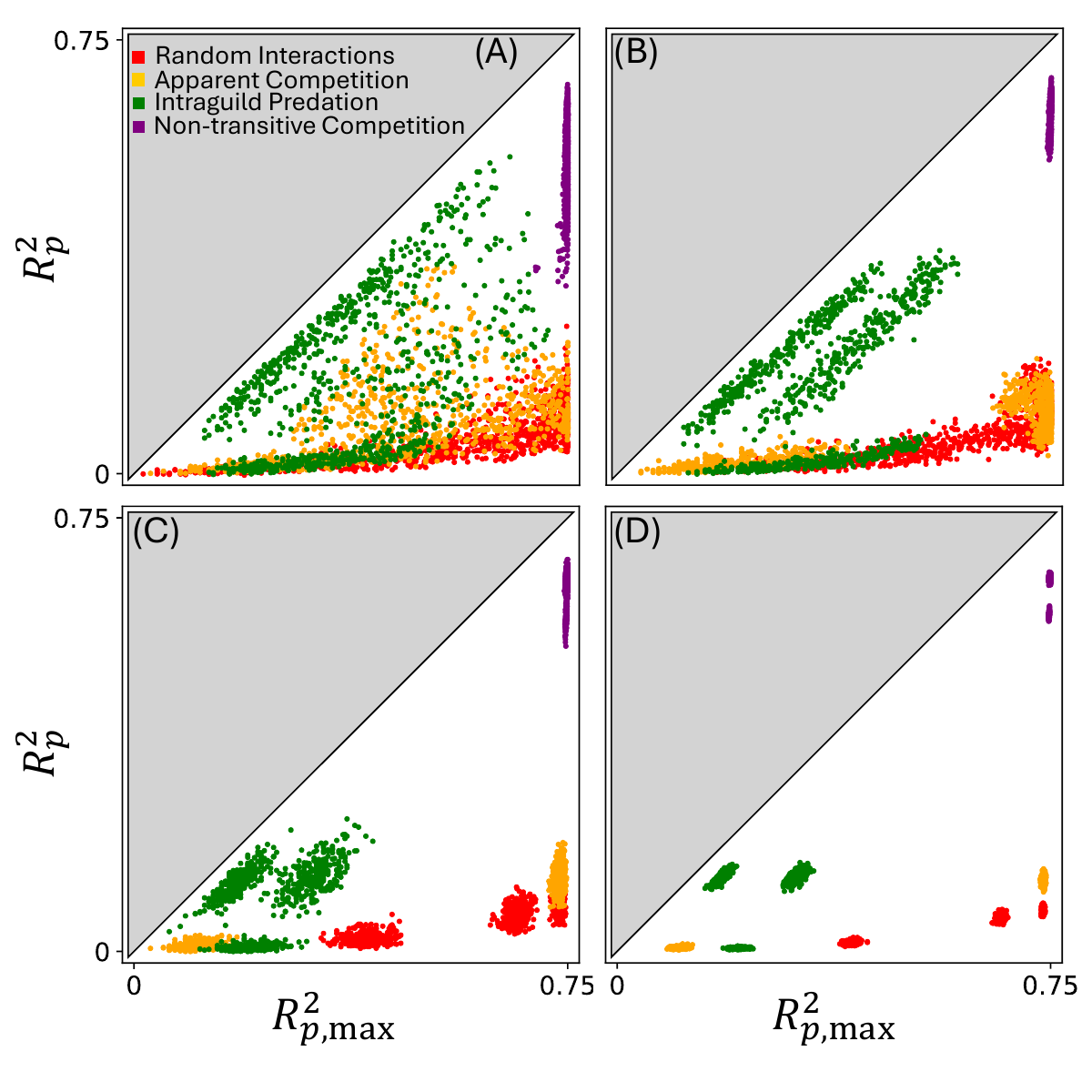}
\caption{A plot of $R_\textrm{p}^2$ versus $R^2_\textrm{p,max}$ for each of the system motifs from the main text. All points are in the lower triangle because $R_\textrm{p}^2 \leq R^2_\textrm{p,max}$, and for logistic regression we have $0 \leq R^2_\textrm{p,max} \leq 0.75$. Each point represents a species-patch combination in a particular simulation. The differences in the spread of the points illustrate the different sources of variability in parameter inference. Panels (A)-(C) are produced from parameter inferences from time series of length $T=1000$. (A) shows a collection of inferences from sets of data, each of which was produced from a new randomization of $\eta_{ij}$ and $\lambda_i$ values. (B) shows the results from simulations in which the diagonal blocks of $\eta_{ij}$ are all identical, but the migration parameters are randomized. (C) shows the results from simulations in which all of the $\eta_{ij}$ and $\lambda_i$ values are identical between simulations, but each point is inferred from a different realization of time series data. (D) is similar to (C) except that the time series used for inference are of length $T=10000$.  }
\label{fig:r2_rmax}
\end{figure}

\begin{figure}
\centering
\includegraphics[width=\textwidth]{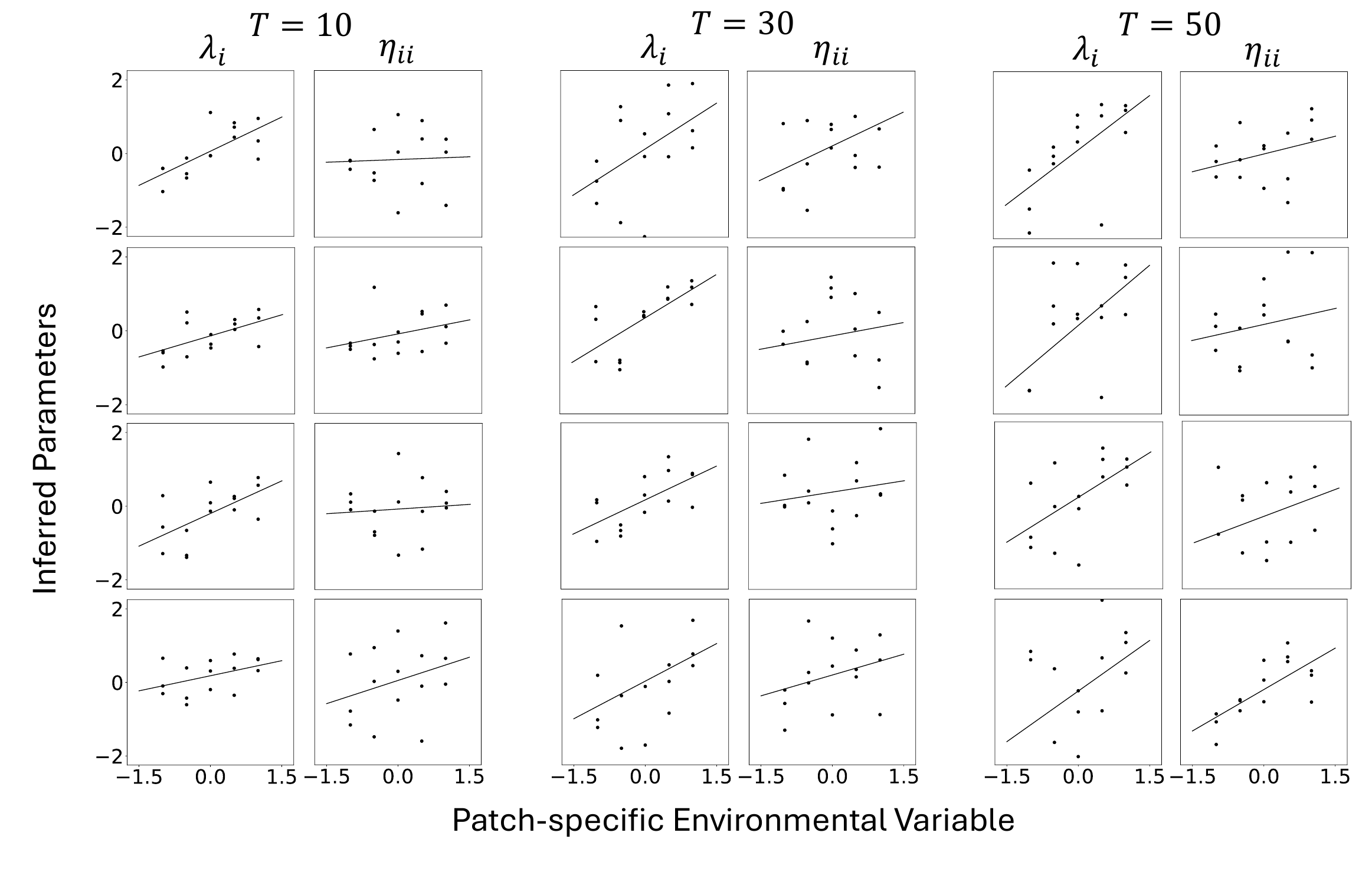}
\caption{Inferring environmental variation from short time series. A metacommunity of five patches and three species with random interactions is simulated. After drawing random parameter values using the same protocol as for Fig.~3 in the main text, the $\lambda_i$'s and the on-diagonal $\eta_{ii}$'s of each patch are added by a random value chosen from $\{ -1, -0.5, 0, 0.5, 1 \}$ (scaled by $0.2$ for $\eta_{ii}$), which represents the environmental suitability of each patch. Simulated times series of different lengths $T$ were used to infer the parameter values, and the inferred values of $\lambda_i$'s and $\eta_{ii}$'s are plotted against the true environmental values added to each entry; different rows are replicate simulations. It can be seen that, even though the estimated value of each parameter is expected to be noisy for such short time series, it is nevertheless possible to detect a significant correlation between the inferred parameters and the environmental variables (black lines are linear regressions). These examples demonstrate that our method can be useful for extracting ecological insight from even short time series data.}
\label{fig:env_inf}
\end{figure}

\end{document}